\documentclass{article}

\usepackage[english]{babel}

\usepackage[hmargin=3cm, vmargin=2cm, includehead]{geometry}

\usepackage{amsmath}
\usepackage{graphicx}
\usepackage[colorlinks=true, allcolors=blue]{hyperref}
\usepackage{lineno}
\usepackage[utf8]{inputenc}
 
\usepackage{amsfonts}
\usepackage{amssymb}
\usepackage{verbatim}
\immediate\write18{texcount -tex -sum  \jobname.tex > \jobname.wordcount.tex}

\title{Automatic Determination of the Weak Beam Condition in Dark Field X-ray Microscopy}

\author{Pin-Hua Huang$^{1}$, Ryan Coffee$^{2}$, Leora Dresselhaus-Marais $^{1,2}$ \\
        \small $^{1}$Department of Materials Science \& Engineering, Stanford University, 476 Lomita Mall, Stanford, CA 94305, USA. \\
        \small $^{2}$SLAC National Accelerator Laboratory, 2575 Sand Hill Rd., Menlo Park, CA 94025, USA. \\
        \small leoradm@stanford.edu
}
\date{} 

\begin{document}

\maketitle
\begin{abstract}
Mechanical properties in crystals are strongly correlated to the arrangement of 1D line defects, termed dislocations. Recently, Dark field X-ray Microscopy (DFXM) has emerged as a new tool to image and interpret dislocations within crystals using multidimensional scans. However, the methods required to reconstruct meaningful dislocation information from high-dimensional DFXM scans are still nascent and require significant manual oversight (i.e. \textit{supervision}). In this work, we present a new relatively unsupervised method that extracts dislocation-specific information (features) from a 3D dataset ($x$, $y$, $\phi$) using Gram-Schmidt orthogonalization to represent the large dataset as an array of 3-component feature vectors for each position, corresponding to the weak-beam conditions and the strong-beam condition. This method offers key opportunities to significantly reduce dataset size while preserving only the crystallographic information that is important for data reconstruction.
\end{abstract}

\section{Introduction}
\begin{figure}[!ht]
\centering
	\includegraphics[width=	129 mm]{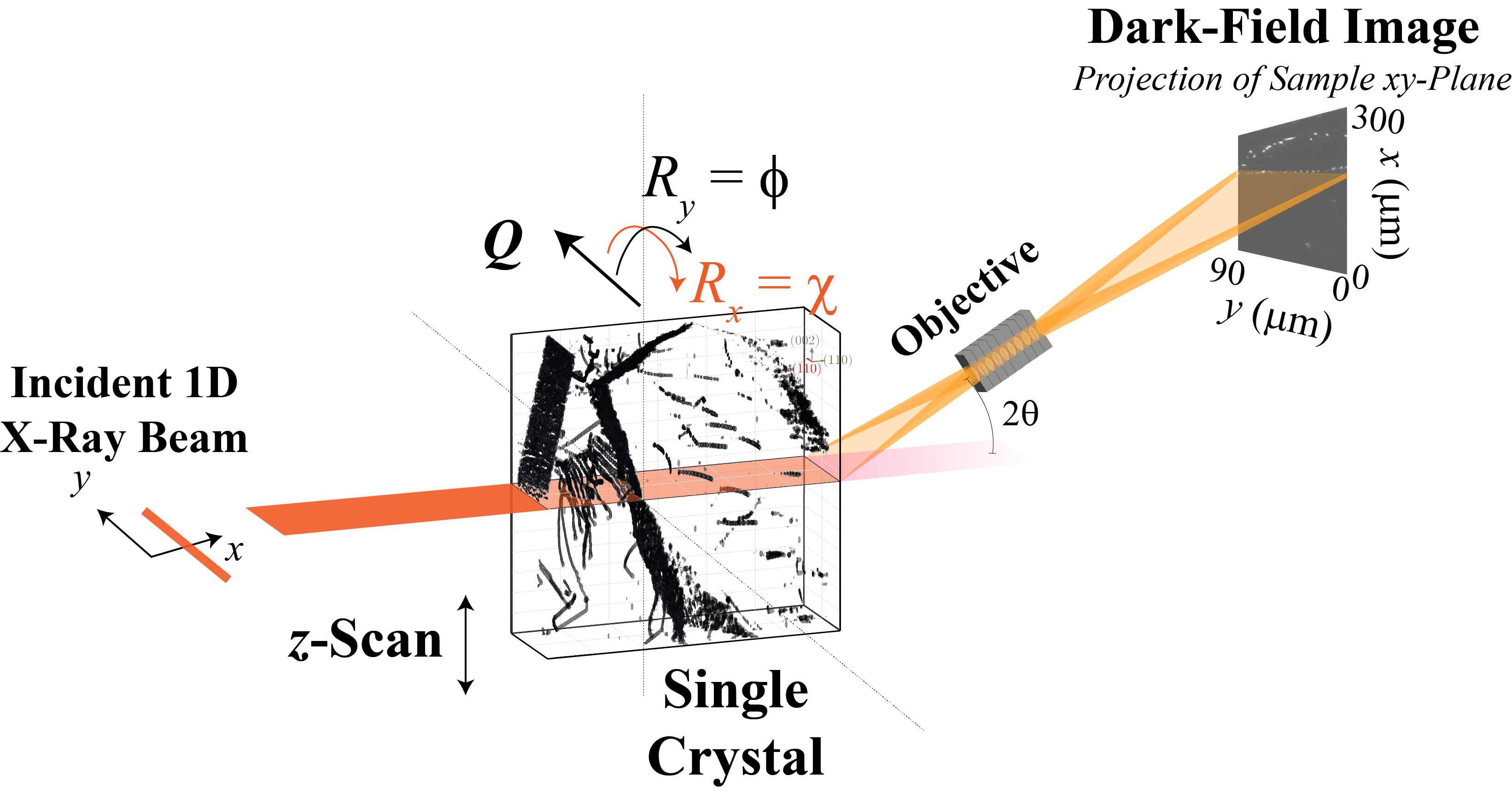}
	    \caption{Schematic showing how a linear x-ray beam illuminates an observation plane in the sample that slices through 3D arrays of dislocations, producing an image with an objective lens along the X-ray diffracted beam. The image shown here is raw DFXM data from this work, showing the weak-beam condition for a series of dislocations. }
        \label{fig:dfxm_setup}
\end{figure}

The mechanical behavior of crystalline materials strongly depends on the arrangement of 1D line defects, termed dislocations. Multiscale characterization of dislocations in bulk materials is essential, but has long been challenging to measurement science. Transmission electron microscopy (TEM) has given us a detailed understanding of dislocation structures in thin films and surfaces of materials, but its subsurface X-ray analogues have lagged in development. Dark-field x-ray microscopy (DFXM) has been developed over the past 10 years to characterize subsurface deformations in crystals by collecting images along an X-ray diffracted beam using an X-ray objective lens (Figure~\ref{fig:dfxm_setup}) \cite{simons2015dfxm}. 

As DFXM images are collected along the diffracted beam, the intensity maps contain information about the material's crystallographic orientation, phase and lattice strain at each position (3D pixel, \textit{voxel}) of the material. 
\textcolor{black}{As 
 such, lattice defects that impose long range distortions within a single grain of the lattice can be observed, so long as they impose asymmetry over the voxel volume ($\sim$70$\times$70$\times600$-nm$^3$). In particular, dislocations (1-D line defects) have cores that impart strain and lattice rotations are on the surrounding crystal that are clearly distinguishable by DFXM, as have been derived elsewhere \cite{poulsen2021}. For sample with sufficiently sparse dislocations, } DFXM has recently been extended to map \textcolor{black}{those} dislocations \cite{jakobsen2019,yildirim2022,porz2021dislocation} and their dynamics \cite{dresselhaus2021}. \textcolor{black}{While imaging dislocations with DFXM is gaining interest, } significant uncertainties persist as to how to characterize them in representative large populations. 

To relate DFXM to materials theory, images must be acquired in large stacks collected from high-dimensional scans to deconvolve the diffraciton signals into maps the long-range elastic distortion fields in each voxel of the material (measured as angular scans through reciprocal space). Full DFXM scans are thus very data-heavy, necessitating data-reduction methods that can extract the information about the strain, rotation, and defect states that represent the physical information relevant to the system. Approaches to develop analysis algorithms to reduce the dimensionality and meaningfully interpret DFXM data have mostly focused on strategies that are general across the vast range of materials systems relevant to DFXM \cite{simons2016dfxm,ferrer2022darfix}; this has focused on assigning the most populated orientations present in each voxel, for each scan axis. This mean-value strategy (called Center of Mass, COM), however, is ill-suited to identify individual dislocations, whose signal is typically better described as weak signal from statistical outliers in orientation and strain \cite{jakobsen2019,yildirim2022}.

The DFXM signal arising from dislocations is superficially analogous to dark-field TEM, in that it defines the locally anomalous packing of a crystal plane. \textcolor{black}{For more direct comparison, we refer to X-ray topography, which also images crystals along a single diffraction-peak in $\vec{q}_{hkl}$. X-ray topography defines the signal in images collected at positions along a rocking curve that samples the Bragg conditions as a function of crystal orientation \cite{Hongyu2021}.} 
An image of a pristine crystal in the most intense Bragg condition shows the entire spatial extent of the undeformed crystal domain, termed the \textit{strong beam} condition. 
Near a dislocation, the crystal has slightly distorted lattice planes emanating from the defect core\textcolor{black}{, generating features that are characteristic of the anomalous crystal packing states, termed the \textit{weak-beam} condition. Dislocations appear characteristically in images collected in the weak beam condition, either as spots or lines, depending on the angle between the observation plane and the dislocation line orientation \cite{Hongyu2021}. The linear arrangement also indicate dislocations piling up in grain boundary. } For a crystal grain with dislocations inside, there are many different weak-beam conditions based on the defects and subdomains inside it, however, there is always exactly one strong-beam condition \cite{jakobsen2019,poulsen2021}. 
\textcolor{black}{Dislocations that slice through the observation plane at steep angles can be }difficult to observe in the strong-beam condition, as the intensity is $\sim$1000$\times$ higher than the weak-beam condition, and \textcolor{black}{the high dynamic range of the images may overwhelm }small shadows.

Recent work demonstrated how supervised statistical methods may be used to map extensive networks of dislocations with DFXM by using dimensional-reduction algorithms \cite{yildirim2022}. The present work extends the previous study by starting to develop a Python library to further automate the computational workflow to identify and segment large populations of dislocations in high-dimensional image stacks. This work focuses on dislocation features in a relatively pristine system: an undeformed and annealed aluminum single crystal, with sample dimensions $0.7 \times 0.7 \times 10$ mm$^3$\textcolor{black}{, as has been described in full elsewhere \cite{yildirim2022}} Our full-field (non-scanning) raw images collected ($x$, $y$) maps of the observation plane in the sample that is illuminated by a 200$\times$0.6 $\mu m^2$ 1D line beam. A full 4D dataset was acquired by collecting images during scans along $z$ (moving up and down) and $\phi$ (rocking scan). This 4D dataset was broken down to understand the relationship between the diffraction condition and the dislocation-relevant features in the image stack. Image processing and feature extracting techniques show that the intensity profiles in certain regions of the image stacks could be used as representative traces that define the training data to define these sets. We use orthogonalization techniques to sort our image stack into three representative component functions that correspond physically to the strong-beam condition, and the two distinct weak beam conditions on either side of the rocking curve. By doing this, our analysis is able to sort difficult-to-interpret features from dynamical diffraction from the weak-beam signals that are characteristic of our dislocation information. Our methods will enable the reduction of data sizes, and will simplify interpretation of DFXM signals to better differentiate the principle features of dislocations for subsequent analysis - even in samples with more than one subdomain.

\section{Experiment Methods}
\subsection{Samples}
The sample we used in this work was a single crystal aluminum of purity 99.99\% with the dimensions of $0.7 \times 0.7 \times 10 $mm$^3$. The samples were used as-purchased from Surface Preparation Laboratories, after being annealed at 590 $^{\circ}$C for 10 hours, then cooled slowly back to room temperature. The sample was kept pristine to retain a low dislocation density in order to capture non-overlapping dislocation features. Further details of this sample are explained in \cite{yildirim2022}.

\subsection{DFXM}
The DFXM experiments were conducted at Beamline ID06-HXM at European Synchrotron Radiation Facility (ESRF) with the same experiment setup as \cite{yildirim2022}.  Figure~\ref{fig:dfxm_setup} shows the setup of the experiment. The line beam produced by 17 keV photon source could illuminate a single plane through the aluminum crystal, which will be captured by an indirect detector. In this paper, we analyzed the images captured in 10$\times$ objectives for a pixel size of 0.75 $\mu$m and a total image view of 2160 $\times$ 2560 pixels. 
This study focuses on dislocation structures observed along the (200) Bragg reflection, at a diffraction angle of \textcolor{black}{$2\theta = 20.77^{\circ}$}. 
The 0.6 $\mu m^2$ 1D line beam produced full-field (non-scanning) images of ($x$,$y$). By scanning along $\phi$, the crystal was rotated through its rocking curve, changing the Bragg condition and corresponding strain-orientation fields that meet the Bragg condition. We collected 31 steps of 0.004$^\circ$ $\phi$ rotations, spanning an angular range of 0.12 $^\circ$ to collect a full set of strong-weak beam conditions in the crystal.
Rocking curves were then collected at a series of 301 layers spaced 1-$\mu$m steps apart, along $z$ (Figure~\ref{fig:dfxm_setup}). \textcolor{black}{The overall dataset volume is around 100 GB of memory for a rocking curve scan which contains [$x$ (2560 pixels), $y$(2160 pixels), $z$(301 layers), $\phi$(31 rocking steps)].} A full 4D dataset mapping the ($x$, $y$, $z$, $\phi$) content of this sample was acquired from the DFXM experiment described in \cite{yildirim2022}.

\section{Data Analysis Method Development}

This study focuses on developing methods to further automate the data reduction for dislocation characterization, beginning with the first \textcolor{black}{$z$-step (which we call a ``rockinglayer'')} from the full 4D dataset. We focus on establishing the relationship between image intensity, $I$, and its parameters $x$, $y$, $\phi$, (i.e. $I(x,y,\phi)$) from DFXM images in the context of dislocation interpretation, using analytical methods viable for computational scaling. 

The primary work we present here is a physics-motivated alternative to principal component analysis. Since the strong-beam (SB) and weak-beam (WB) conditions  are known to represent the physics of interest for this system \cite{Cockayne1971}, instead of using purely data-science approaches to decompose the 3D dataset into arbitrary principal components, we focus on defining the components that are physically meaningful to our system scientifically, then decompose the \textcolor{black}{dataset} into only those components. While this approach is less general, it provides the results most meaningful to dislocation science.

\textcolor{black}{We first define a method to reduce our 31-\textcolor{black}{rocking }step $\phi$-resolved scan of the same positions in terms of its three meaningful components that describe the SB, WB$_-$ ($\phi < \phi_{SB}$) and WB$_+$ ($\phi > \phi_{SB}$) components present. To do this, we define all three components as orthogonal principal vectors.} The SB component describes the crystal domain's center orientation, as is computed in the conventional analysis, while the WB contrast shows ``defect bands'' in reciprocal space that \textcolor{black}{give information about the} dislocations. \textcolor{black}{WB$_-$ and WB$_+$ display signal from the anomalous components of $q$ that appear before and after the rocking steps of the SB components, respectively, giving information relevant to the dislocations present.}
In dislocation studies, identification of the dislocation core's line vector and \textcolor{black}{Burgers vector} (the amplitude and direction of crystallographic shear the dislocation imparts onto the lattice) are required to characterize.
For DFXM signal, the WB components map the positive and negative displacement (related to strain and rotation) components of each dislocation. This means that the ``true'' position of a dislocation core actually lies at an ($x,y$) position between the two spatial components of WB$_+$ and WB$_-$, necessitating that data reduction algorithms capture the information contained on both sides of the rocking curve. 

\begin{figure}[!ht]
\centering
	\includegraphics[width=129 mm]{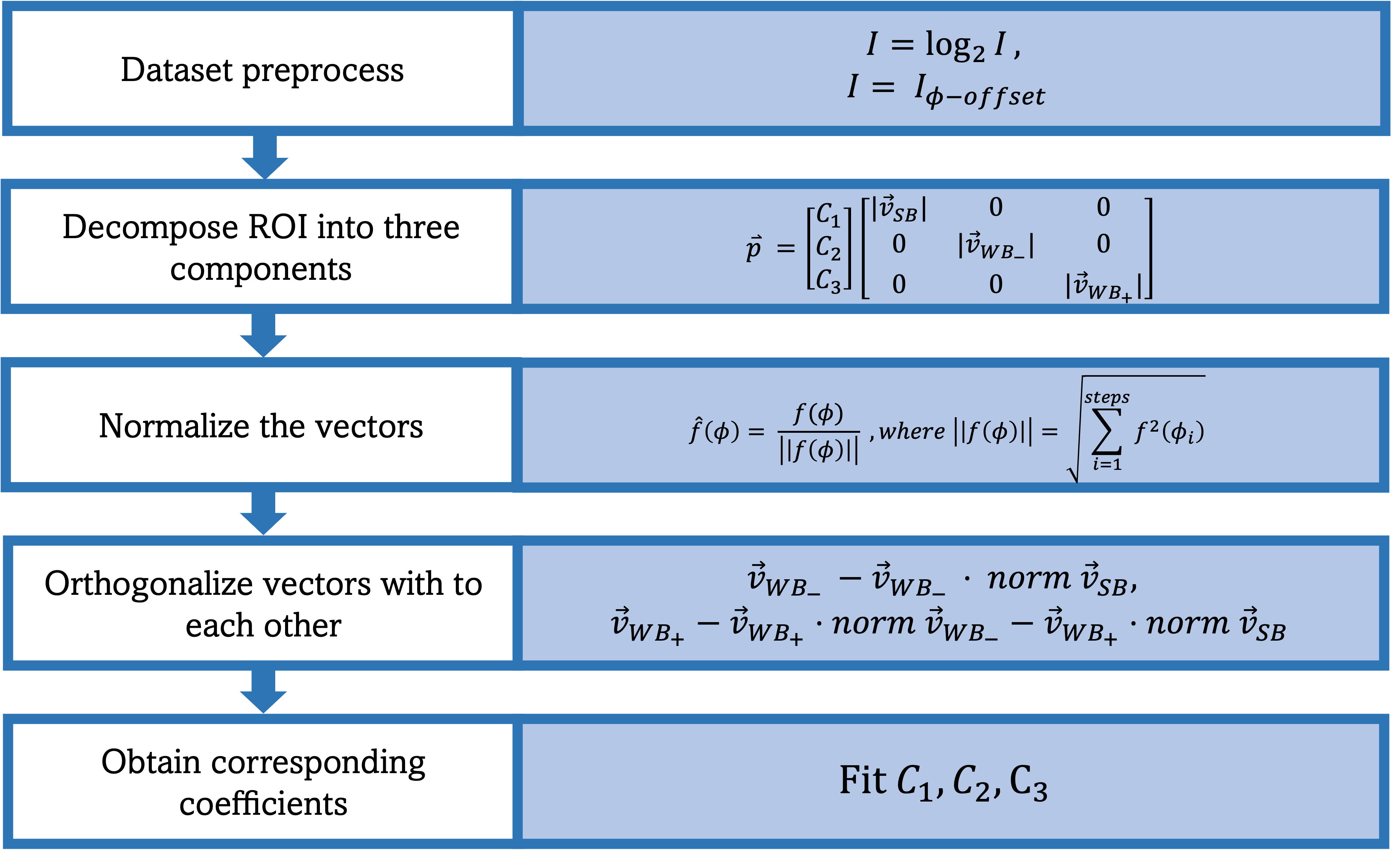}
	    \caption{Flow chart of the steps required for Gram-Schmidt data analysis.}
        \label{fig:flow_chart}
\end{figure}

As such, we elected to reduce the full 3D dataset from the $2560\times2160\times31$-pixel$^3$ dataset into a stack of three 2D images to describing the three physical components described above.\textcolor{black}{ We outline the workflow for our data analysis procedure in Figure~\ref{fig:flow_chart}. Our code from this work is available on Github at \url{https://github.com/leoradm/DFXM_GramSchmidt}.}

\subsection{Pre-processing and ROI selection}

\textcolor{black}{For the Gram-Schmidt procedure to work effectively in this case, images requires some pre-processing to effectively distinguish the representation regions of interests for each principle component. To do this, we first read the 3D dataset ($x, y, \phi$) into Python as a Numpy array from the  *.edf files acquired at the ESRF using Python Darfix library \cite{ferrer2022darfix}.
The dynamic range of the original dataset is then converted from the original $I(\phi)$ to its log base 2 version to amplify the dislocation features present in the low-intensity regions of the image, for best visualization, i.e., }
\begin{equation}
     I'_{x,y} = log_2(I_{x,y}).
    \label{eq:takelog}
\end{equation}
\textcolor{black}{To define and orthogonalize the three characteristic vectors, we needed to remove the crystallographic offsets between different subdomains (caused by the dislocation boundaries). We emphasize that this ``correction'' is crystallographic information that we remove in this case because the misorientation between subdomains is already identifiable by the COM techniques developed previously \cite{ferrer2022darfix}. Thus we remove this information from our WB-focused analysis, we computed the pixel intensity $I_x,y$ for center of mass in terms of rocking steps ($\phi$) is calculated as }
\begin{equation}
    I''_{x,y} = I'_{x,y} - (I'_{x,y})_{0.01},     
    \label{eq:colormap1}
\end{equation}
\textcolor{black}{where $(I'_{x,y})_{0.01}$ describes the lowest 1\% intensity values that remove the baseline intensity. We define the colormap (shown in Fig.~\ref{fig:COM}a) as }
\begin{equation}
    \textit{colormap }I_{x,y} = \sum_{n}^{s}I_{x,y}\times n \text{, s = 31}
    \label{eq:colormap2}
\end{equation}
\textcolor{black}{where $n$ is the rocking step and $s$ is the total rocking step number. We then blurred the colormap using $9\times9$ pixel range to generalize the crystallographic orientation around each pixel, as}
\begin{equation}
    \textit{blurred colormap }I_{x,y} = \frac{\sum_{x=-4}^{4}\sum_{y=-4}^{4}I_{x,y}}{9\times9}
    \label{eq:colormap3}
\end{equation}
\textcolor{black}{The colormap is shown in Figure~\ref{fig:COM}.
To establish the offset correction, we then define a threshold by manually selecting the number that best separates two major crystallographic domains, in this case, $I_{thresh} = 10$, as shown in Fig.~\ref{fig:COM} with }
\begin{equation}
    \begin{gathered}
    \textit{if } I_{x,y}<I_{thres}\text{, }I_{x,y} = 0\\
    \textit{if } I_{x,y}\geq I_{thres}\text{, }I_{x,y} = 1
    \end{gathered}
\end{equation}
\textcolor{black}{To separate the dislocation information within the thresholded region, we used \textit{findContours()} contouring method from the OpenCV library in Python to keep the large scale half side of the offset. This method thresholds the image based on the second derivatives of the pixel intensity, }
\begin{equation}
\begin{gathered}
    \textit{contour method: if }\frac{\partial I_{x,y}}{\partial x}\frac{\partial I_{x,y}}{\partial y} = 0\text{, }I_{x,y} = 0\\
    A\geq A_{0.995} = 1\\
    A<A_{0.995} = 0
\end{gathered}    
\end{equation}
\textcolor{black}{the offset is corrected by shifting the intensity value by the offset value, which is one rocking step difference here, along $\phi$ value}
\begin{equation}
    I_{\phi} = I_{\phi+offset}
\end{equation}

\begin{figure}[!ht]
\centering
	\includegraphics[width=129 mm]{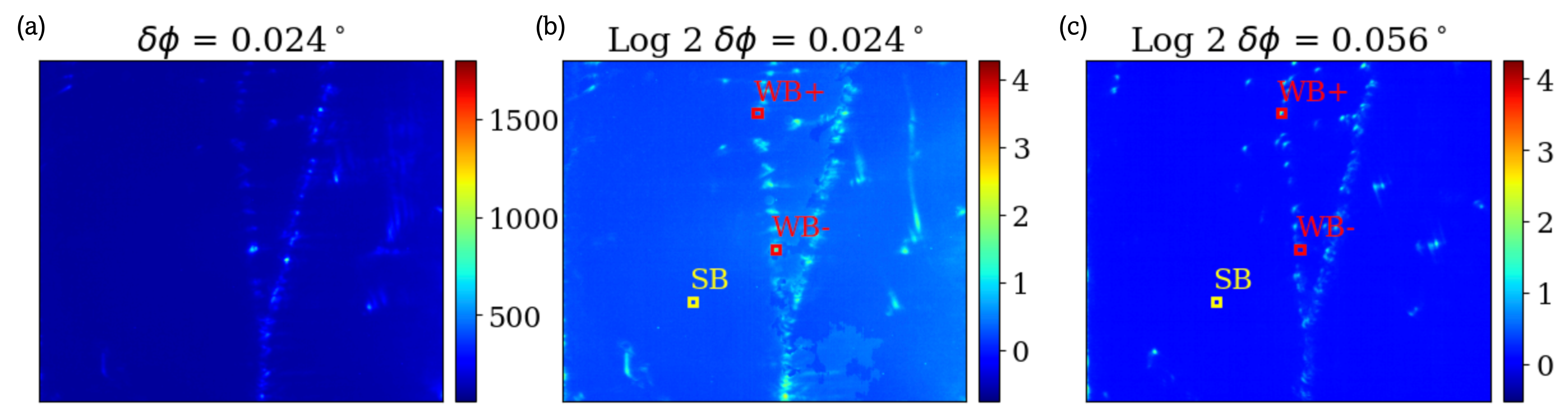}
	    \caption{Selected DFXM frames (plotted in pseudocolor for clarity) at rocking positions of (a) $\phi$ = 0.024$^\circ$ for original dataset, (b) log 2 base $\phi$ = 0.024$^\circ$ to show the Weak-Beam Left condition, and (c) log 2 base $\phi$ = 0.056$^\circ$ to show the Weak-Beam Right conditions. \textcolor{black}{Strong beam region of interest in black box is used to define a basis function of the unperturbed background crystal matrix and the weak beam regions are plotted in the red boxes.}} 
        \label{fig:training_data}
\end{figure}

With the pre-treated dataset, the image was then divided into ($x,y,\phi$) subarrays that were used to define each basis function, and to define a smaller region of interest (ROI) for computational efficiency, as shown with the labelled boxes in Figure ~\ref{fig:training_data}.

\subsection{Defining Orthogonal Basis Functions}

We selected three 50$\times$50-pixel$^2$ regions to specify the $I(\phi)$ functions most representative of the SB, the WB$_+$ and the WB$_-$ conditions. 
The region selected for the strong-beam condition (labelled ``SB'' in \textcolor{black}{black} in Figure~\ref{fig:training_data}) contained no visible dislocation features for the entire $\phi$ extent of the rocking curve. 
To ensure the information contained in each WB condition corresponded to defects of interest, we \textcolor{black}{manually} identified the frame representative of the weak-beam condition on each side of the rocking curve, and identified dislocations whose $I(\phi)$ were most selectively populated on only one side of the rocking curve. 
As shown by Figure~\ref{fig:training_data}, some signal was still present on the other side of the rocking curve, requiring some data treatment afterwards for full \textcolor{black}{orthogonality} of the training functions. 

The characteristic regions defined in Figure~\ref{fig:training_data} were then used to define the orthogonal set of basis functions to uniquely decompose the dataset. 
By defining normalized basis functions, we \textcolor{black}{proposed that} our intensity function at each pixel $p_{x_i,y_j}(\phi)$ \textcolor{black}{could be divided} into the three corresponding \textcolor{black}{principle} components, \textcolor{black}{$\hat{v}_{\text{SB}}$, $\hat{v}_{\text{WB}_-}$, and $\hat{v}_{\text{WB}_+}$}, and decompose the image into arrays of the scaling coefficients \textcolor{black}{$c_1$, $c_2$, $c_3$} to display each type of information uniquely. 
\textcolor{black}{
\begin{gather}
 p_{x_i,y_j}(\phi)
 =
  \begin{bmatrix}
   c_1(\phi)\\
   c_2(\phi)\\
   c_3(\phi) \end{bmatrix}
  \begin{bmatrix}
   \hat{v}_{\text{SB}} & 0 & 0 \\
   0 & \hat{v}_{\text{WB}_-} & 0\\
   0 & 0 & \hat{v}_{\text{WB}_+}\end{bmatrix}
   \label{eq:decompose_matrix}
\end{gather}}
To define the functions $\hat{v}_{\text{SB}}$, $\hat{v}_{\text{WB}_+}$ and $\hat{v}_{\text{WB}_-}$, each $50\times50$-pixel$^2$ region was converted to the un-normalized $\Vec{v}_{i}$ by calculating the mean intensity across all pixels at each value of $\phi$, 
\begin{equation}
    \Vec{v}(\phi)=\frac{\sum_x^{50}\sum_y^{50}I_{x,y,\phi}}{50^2}.
    \label{eq:mean_intensity}
\end{equation}
\textcolor{black}{To orthogonalize the principle components, we begin with normalizing SB condition following the Gram-Schmidt process,} 
\begin{equation}
    \hat{v}_{\text{SB}}(\phi) = \frac{\vec{v}_{\text{SB}}(\phi)}{||\vec{v}_{\text{SB}}(\phi)||}, \hspace{2em} \text{where } ||\vec{v}_{\text{SB}}(\phi)|| = \sqrt{\sum^{steps}_{j = 1}\vec{v}_{\text{SB}}^2(\phi_i)}.
    \label{eq:normalize_SB}
\end{equation}
\textcolor{black}{The first normalized principle component, $\hat{v}_{\text{SB}}$, is orthogonalized and removed from the two WB components started from WB$_{-}$
\begin{equation}
    \vec{v}_{\text{WB}_-} = \vec{v}_{\text{WB}_-} - \vec{v}_{\text{WB}_-}\cdot proj_{\vec{v}_{\text{WB}_-}}(\hat{v}_{\text{SB}})
    \label{eq:GS_WB-}
\end{equation}
}
\textcolor{black}{
The WB$_-$ principle component was then obtained by normalizing
the resulting $v_{\text{WB}_-}$,}
\textcolor{black}{
\begin{equation}
    \hat{v}_{WB_-}(\phi) = \frac{\vec{v}_{WB_-}(\phi)}{||\vec{v}_{WB_-}(\phi)||}, \hspace{2em} \text{where } ||\vec{v}_{WB_-}(\phi)|| = \sqrt{\sum^{steps}_{j = 1}\vec{v}_{WB_-}^2(\phi_i)}.
    \label{eq:normalize_WB-}
\end{equation}
}
\textcolor{black}{Finally, the remaining WB$_+$ principle component is obtained by removing its projection onto both the SB and the WB$_-$, then normalizing the result,
\begin{equation}
\begin{gathered}
      \vec{v}_{\text{WB}_+} = \vec{v}_{\text{WB}_+} - \vec{v}_{\text{WB}_+}\cdot proj_{\vec{v}_{\text{WB}_+}}(\hat{v}_{\text{SB}}) - \vec{v}_{\text{WB}_+}\cdot proj_{\vec{v}_{\text{WB}_+}}(\hat{v}_{\text{WB}_-}) \\
      \hat{v}_{WB_+}(\phi) = \frac{\vec{v}_{WB_+}(\phi)}{||\vec{v}_{WB_+}(\phi)||}, \hspace{2em} \text{where } ||\vec{v}_{WB_+}(\phi)|| = \sqrt{\sum^{steps}_{j = 1}\vec{v}_{WB_+}^2(\phi_i)}.
    \label{eq:GS_WB+}  
\end{gathered}
\end{equation}
}

\textcolor{black}{As shown above, the principle components are obtained by removing projection of other principle components and normalizing the vectors. In this way, our full image stack may be describe based on the coefficients to describe the components of all three principle components, defined in the space, \{$\hat{v}_{SB}, \hat{v}_{WB_-}, \hat{v}_{WB_+}$\}.}

\subsection{Image Decomposition}
Once the complete set of orthogonal basis functions were defined for our dataset, the components of each function could be computed uniquely to decompose the image into its grain and defect features. To do this, we decomposed the image and computed its coefficient by multiplying each pixel's intensity function $p_{x_i,y_j}$ with each of the 3 basis functions, $\hat{v}_{\text{SB}}$, $\hat{v}_{\text{WB}_+}$, and $\hat{v}_{\text{WB}_-}$. Three 2160 $\cdot$ 2560 pixel$^2$ images is acquired by summing through the rocking steps.
The ROI used to demonstrate this approach is shown by the boxes on Figure~\ref{fig:training_data}. 

\section{Results}

\begin{figure}[!ht]
\centering
	\includegraphics[width= 89 mm]{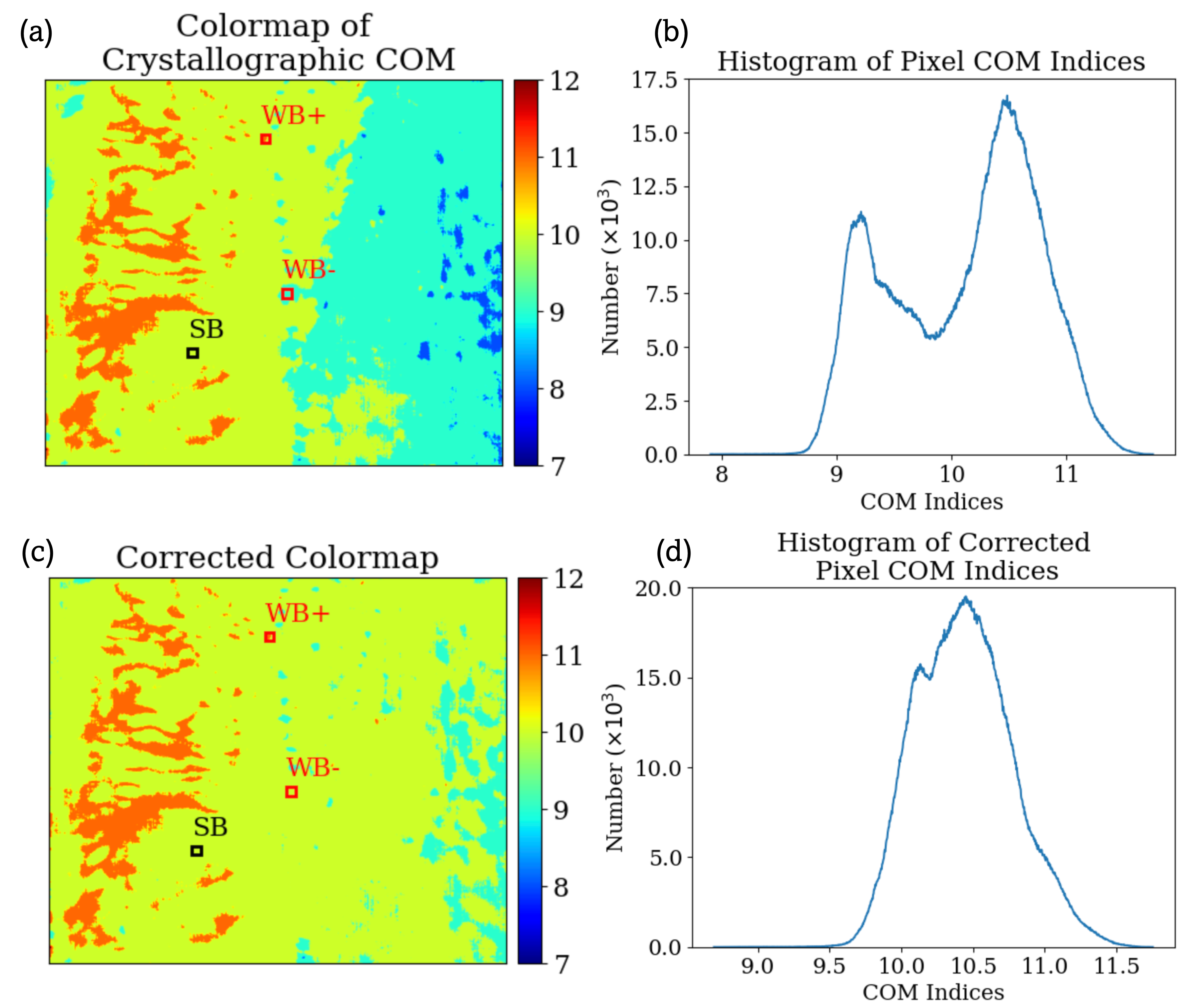}
	    \caption{(a) Original colormap to show crystallographic orientations of each pixel and (b) the histogram of the pixel COM value.  (c) The crystallographic offset corrected colormap and (d) the corrected histogram of the pixel COM value}
        \label{fig:COM}
\end{figure}

\textcolor{black}{We display the results of our initial COM analysis and the offset results from our pre-processing treatment in Figure~\ref{fig:COM}. The intensity throughout rocking step $I(\phi)$ includes the crystal orientation with the highest intensity in the rocking scan for each pixel, implying the most prevalent local crystal orientation. In this case, we represent the most likely orientation using the rocking-step for clarity. 
The bimodal nature of our distribution is evident in the histogram we show in Fig.~\ref{fig:COM}b, which indicates the COM of all pixels in the image, based on the un-rounded results from Equation \ref{eq:colormap3}. The two peaks of the histogram's distribution are at step 9.2 and 10.7 which corresponds to $\phi=0.0368^{\circ}$ and $\phi=0.0428$ in theoretical, respectively. However, they are imaginary numbers and based on the discrete rocking step ($0.004^{\circ}$), we could only increase the COM index by integer multiplication of a single rocking step. Based on the values from this histogram, we thus selected $I_{\text{thresh}}=10$ to describe the cutoff between distributions, and corrected all pixels for $I_{\text{COM}}$ less than $I_{\text{COM}}=10$ distribution to increase their $\phi$ indexes by $0.004^{\circ}$ to match the upper distribution, as shown in the COM image in Fig.~\ref{fig:COM}c and associated Fig.~\ref{fig:COM}d histogram. 
In order not to lose the dislocation information embedded in the teal color side, we implemented the algorithm to exclude dislocation island information around the boundaries and only keep the large right side of it. For pixels with COM index values larger than 10, $p_{x,y}(\phi)$ functions were shifted one index back to align to the nearest Bragg condition. The corrected result is shown in Figure~\ref{fig:COM}c and d.}

\textcolor{black}{Studies shown that dislocations pack into low angle boundaries that misorient the crystal across the boundary or pile-up \cite{Vogel}. While our preprocessing treatment removes this information, we note that the original COM data still retains it if necessary to the analysis. }

\textcolor{black}{After offset correction, the colormap shows no significant difference in crystallographic orientation difference and the histogram in Fig.~\ref{fig:COM}d shows that most of COM of the pixels lay in the range between 10 ad 11. This treatment prepares the images to ensure that the three characteristic SB, WB$_+$ and WB$_-$ define the complete set required to describe the image stack. While there are still some remaining fringes in Figure~\ref{fig:COM}c, they are characteristic of the dynamical diffraction, which is characteristic of the SB condition's dynamical diffraction image features \cite{saleh2019fundamentals}}. 

\begin{figure}[!ht]
\centering
	\includegraphics[width= 129 mm]{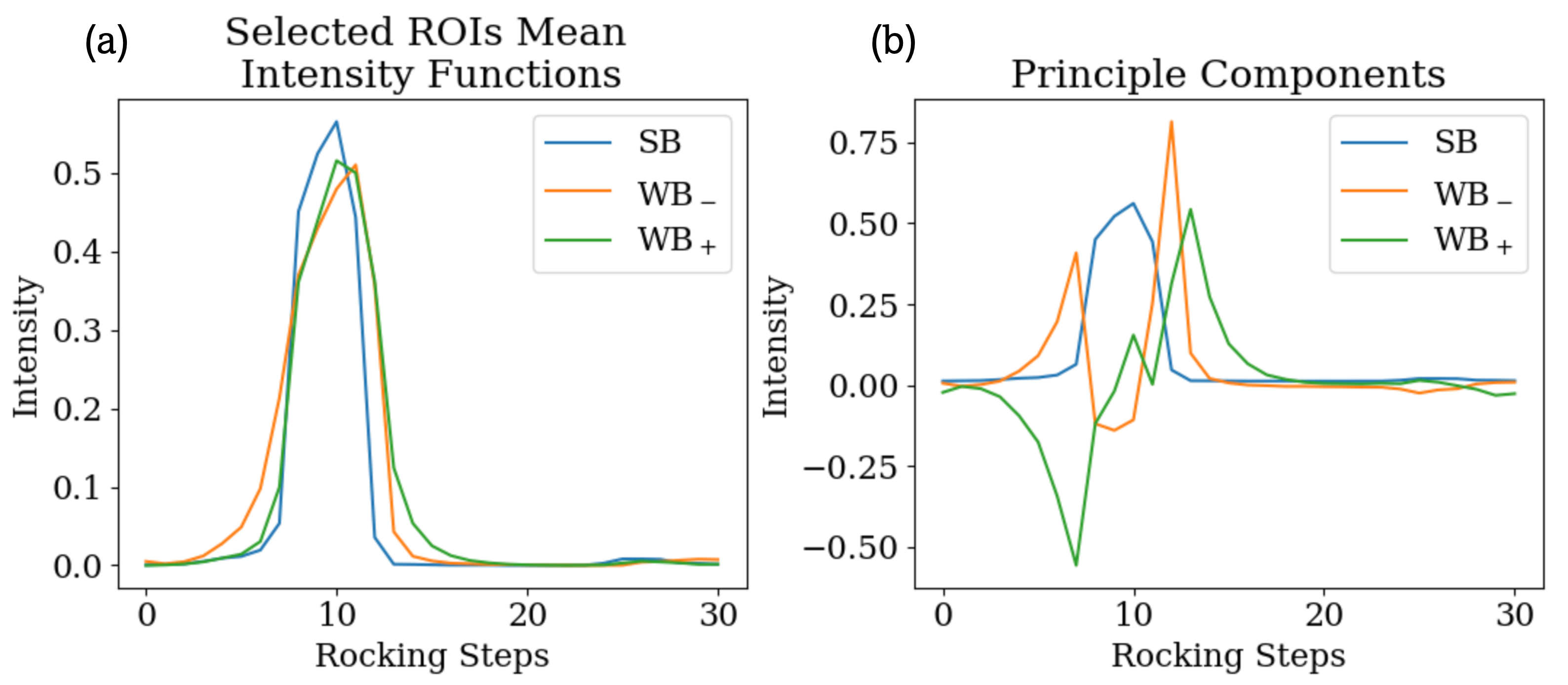}
	    \caption{(a) Normalized mean intensity along rocking steps of 3 selected ROIs (b) Principle components of SB, WB$_-$, and WB$_+$.}
        \label{fig:GS}
\end{figure}

In Figure~\ref{fig:training_data}, we show traces for the manually-defined regions that were characteristic of $\hat{v}_{\text{SB}}$, and the positive and negative components of the rocking-curve edges, $\hat{v}_{\text{WB}_-}$ and $\hat{v}_{\text{WB}_+}$, respectively. 
\textcolor{black}{The raw intensity collected from each training data, $I(\phi)$, is plotted in Figure~\ref{fig:GS}a for the unnormalized functions to display the anomalous distortions that characterize each function utilizing Equation~\ref{eq:normalize_SB}. 
Figure~\ref{fig:GS}b shows the selected principle components and that after being orthogonalized. }

\begin{figure}[!ht]
\centering
	\includegraphics[width= 129 mm]{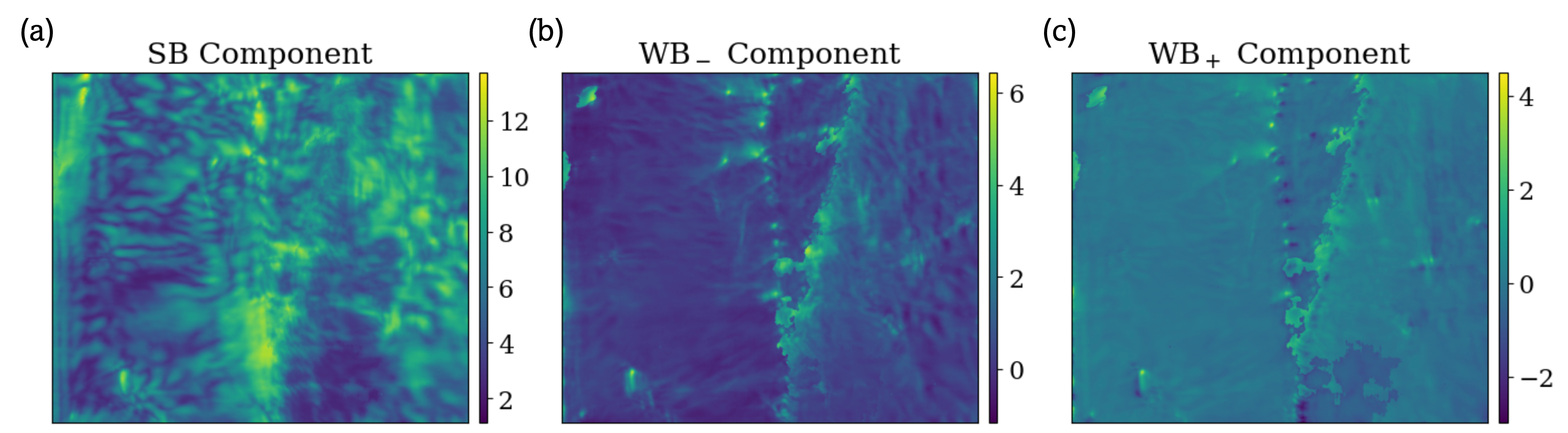}
	    \caption{(a) SB component, (b) WB left component and (c) WB right component of decomposed dataset}
        \label{fig:results}
\end{figure}

\textcolor{black}{Based on the three orthogonal components described in Fig.~\ref{fig:GS}b, the coefficients to decompose the image into its principle components spatially map the coefficient images plotted in Figure~\ref{fig:results}.}

\textcolor{black}{The majority of the dynamical diffraction fringes are present in the SB component (Fig.~\ref{fig:results}a). By contrast, the characteristic information about the distortions (in this case dislocations) are present in the two WB components. In the WB$_-$ component (Fig.~\ref{fig:results}b), the features in dislocation structures have higher intensity than the background intensity. By comparison, the WB$_+$ components (Fig.~\ref{fig:results}c) are mostly low-intensity features with a constant-intensity background. }

\section{Discussion}
\textcolor{black}{The three principal components to describe the dislocation information in our DFXM dataset are the SB information (Fig.~\ref{fig:results}a), the WB$_-$ (Fig.~\ref{fig:results}b), and the WB$_+$ (Fig.~\ref{fig:results}c). We note that the features arising from dynamical diffraction (multiple scattering in XRD that generates Pendell\"{o}sung fringes) should only occur in regions with very strong light-matter interactions, and therefore should only be observed in the SB component of our result. By contrast, the WB components we resolve have significantly lower signal, and therefore should resolve signal that obeys kinematic diffraction (without fringes) because of its much lower scattering intensity. As the SB signal we separate is 160$\times$ higher in counts than the WB$_-$ and WB$_+$ signals and contains almost all the fringed features, our approach effectively decomposes the present dataset.}

\textcolor{black}{While dislocation features are most evident in the weak-beam condition, the rotation imposed by dislocation boundaries often causes DFXM images to contain information from both weak-beam and strong-beam signals in different regions of the material. }
In a single-grain sample, the use of only one set of SB, WB$_+$ and WB$_-$ may be sufficient to describe all WB data in the set, however, this is usually not the case. 
As DFXM has extremely sensitivity to lattice distortions, the 3D volume of most crystals includes multiple SB and WB conditions that complicate the interpretation of dislocations due to parasitic changes to the detector's sensitivity to WB signal when spanning the full dynamic range of intensities from $\sim$150 to $\sim$6000 readout counts. Manually selected WB frames are thus often not representative of the full duration of the scan, or could represent anomalous defect features that are not representative of the full population in the 3D volume. In this case, our selections provided a means to refine and demonstrate the viability of our approach to decompose the full dataset into its 3 physically meaningful values required for subsequent materials-science analysis. 

In this work, dislocation features were only explored in the $I(x,y,\phi)$ 3D subset of the data, \textcolor{black}{focusing on automating the WB-identification step that was described in our previous study \cite{yildirim2022} to enable subsequent dimensional reduction methods to identify the dislocations (e.g. Fig.~\ref{fig:dfxm_setup}).}
In future, we plan to expand the present approach to the full 4D datasets $I(x,y,z,\phi)$ \textcolor{black}{for $z$-resolved experiments that study more intricate networks of crystal subdomains that may start and terminate at different positions through the height of the crystal. 
For the long-term impact of this approach, we that our approach to pre-processing that makes the program robust to the different COM orientations will need to be fully automated}.
For our 3D imaging goals, the ``pixel'' intensities must be expressed as 3D ``voxel'' intensities, requiring 3D processing methods in place of 2D kernels and pixelation discussed in this work. 
In this particular dataset, the $z$-layers were spaced 1 $\mu$m apart, introducing $75\times200\times1000$-nm voxel sizes. 
This implies that reconstruction of 4D dataset has significantly less neighborhood information and associated resolution in the $z$ direction, making that dimension's pixelation possibly less reliable. Despite its challenges, this approach is absolutely worthwhile, as it constrains the dislocation character and structure required to inform the crystal's mechanical and physical properties \cite{yildirim2022}.
\textcolor{black}{We also see opportunities to further automation these methods, including method to automatically identify the basis functions, $\hat{v}_i$, for on-the-fly data analysis during experiments that could help guide the science in real-time for \textit{in-situ} or \textit{operando} experiments.}

\textcolor{black}{We note that this work has focused on studies for pristine material samples, which are very relevant to semiconductor and ceramic systems. For higher dislocation density systems with spaced dislocation boundaries, smaller modifications to the training functions would enable this analysis to sort the principal components based on dislocation structures vs the undeformed crystal. By contrast, studies of high dislocation density systems upon significant deformations would likely require other methods to account for all orientations of the different subdomains, as described in \cite{yildirim2022snake}. }

\section{Conclusion and Outlook}
This work introduces a new computational method to identify and characterize information unique to dislocations from the low-intensity weak-beam information in 3D data obtained by DFXM. We demonstrate that by compiling 3 orthogonal basis functions to describe the SB, WB$_+$ and WB$_-$ conditions, we can decomposing the 3D dataset into its different defect and grain components.
Using this approach, we reduce the initially 31-dimensional $\phi$-resolved data into 3 images, whose principal components directly map the dislocation and grain components in our crystal. 
Our results show that this orthogonalization and decomposition allows us to customize our dimensional reduction algorithms for physically meaningful data specific to dislocation theory. This approach will be especially useful in interpreting $\geq$4D scans (e.g. Figure~\ref{fig:dfxm_setup}). Our future work will extend this approach to higher dimensions, aided by new machine-learning tools for fully automated and unsupervised versions of this data reduction.

\section*{Acknowledgements}
We acknowledge the European Synchrotron Radiation Facility (ESRF) for provision of synchrotron radiation facilities and we would like to thank Can Yildirim for conducting the experiments at beamline ID06-HXM.

\section*{Declarations}

 On behalf of all authors, the corresponding author states that there is no conflict of interest.

\bibliography{reference}
\end{document}